# Synthesis and characterization of Zn doped Mn ferrites nanostructures

**Saima Rani[1] and Syed Shahbaz Ali[1,*]**

**Department of Physics, The University of Lahore, Lahore Pakistan**

## Abstract

Zn doped Mn ferrites nanoparticles were fabricated by using Co-precipitation. Variation in structure, magnetic and optical properties of MnZn ferrites has been discussed. First of all, samples were synthesized, annealed at (400°C, 500°C, 600°C & 700°C) and then characterized. The as-synthesized and annealed samples were investigated by X-ray diffraction (XRD), Scanning electron microscopy (SEM), Energy Dispersive spectroscopy (EDX), Ultra Violet visible spectrometry (Uv-Vis spectrometry) and Vibrating sample magnetometer (VSM). The average crystallite size of MnZn ferrites nanoparticles determined from XRD were in the range of 42 to 60 nm. These nanoparticles possess normal spinel structure. The SEM images showed the physical shape of the samples, which showed that the as prepared samples are more agglomerated and having flake like shape rather than annealed at 700ºC while the samples have longitudinal or rod like shape on annealing at 700ºC. The coercivity (Hc), saturation magnetization (Ms), and remanence (Mr) of Np's were also calculated. The (Ms) value is increasing from 26 to 65 emu/g, the coercivity (Hc) is varying from 13 to 193 Oe and remanence (Mr) has also showing increasing trend although very less, from 0.031 to 0.798 emu/g which are a little part of their bulk counter parts. The band gap energy of the samples was showing decreasing trend as with the increase of particle size which is of the order of 3.5 to 2.9 eV.

## 1. Introduction

A lot of substitutes have been added in ferrites ($Fe_2O_4$) as doping to enhance the magnetic and electric properties of ferrites. These ferrites can be prepared in different forms like in thin films or in powder form. A large number of methods have been developed to synthesize ferrites; some of them are the hydrothermal process, the co-precipitation method, the micro-emulsion method, and sol gel synthesis. After the study of literature about soft ferrites it has been cleared that chemical co-precipitation method is much easier, low cost, and convenient to control particle size and purity. Soft ferrites are a type of ferrites having superior magnetic properties used in electrical and microwave industry. MnZn ferrites got a deep interest for research from researchers' just because of application point of view in previous years. Its preparation method, composition, doping, calcination or annealing changed its properties and application. That why it's a deep ocean of properties variation. Soft ferries are used in stem up, step down transformers, in recording heads, in remote devices etc.

C. Venkataraju et al investigated the cation distribution effects on the structural and magnetic properties of Nano particles of $Mn_{(0.5-x)}Ni_xZn_{0.5}Fe_2O_4$ (where x varied from 0.0, 0.1, 0.2, 0.3) prepared by Co-precipitation method [1]. C.F. Zhang et al investigated the structural and magnetic properties of MZn ferrite NP's with different doping values of cobalt by Co-precipitation method [2]. Darko Makovec et al studied the spinel structure of MnZn ferrite nanoparticles by Co-precipitation in reverse micro-emulsion method [3].

Xie Chao et al studied the effect of pH on Mn–Zn ferrites properties synthesized from low grade manganese ore (LMO). He prepared Mn-Zn ferrites powder from LMO by co-precipitation combining with ceramic preparation method [4]. N. D. Kandpal et al studied the synthesis and characterization of Iron oxide with the help of cost effective co-precipitation method. Different techniques like XRD, TEM, VSM and FT-IR were followed to identify the lattice parameters, crystal structure and the magnetic properties [5].

## 2. Experimental

**Synthesis and annealing process of $Mn_{0.5}Zn_{0.5}Fe_2O_4$**

For the preparation of Zn doped Mn ($MnZnFe_2O_4$) ferrites nanoparticles by Co-Precipitation. We used $Fe2O_4$, ZnCl2 and MnCl2 as the starting precursors. The 0.5 M aqueous solution of each salt was prepared by adding these salts in 50 ml of distilled water. NaOH worked as precipitating agent, its 0.64 M aqueous solution was also prepared. All these three initially prepared precursor solutions were mixed by using magnetic stirrer continuously. The 0.5 M aqueous solution of each salt was fabricated to get required atomic ratio of final product. During the mixing of three initial precursor solutions, NaOH was added drop wise to the precipitation to control the pH of the solution. During the mixing process a constant temperature of 80°C was provided continuously for 2 hours for better results of reaction process. After sufficient precipitation has been observed, the reaction was stopped and the precipitates were allowed to settle down at the bottom of beaker. Later on the liquid was taken out by using a simple sucker and the precipitates was washed 4 to 5 times with distilled water and then at least 2 times with ethanol. The precipitates were dried out for about 4-6 hours in a drying oven at a temperature of about 80°C and then the final product was grinded by using mortar and pestle to get uniform powder which contained fine nanoparticle of our desired materials.

Following reaction took place.

$$MnCl_2.3H_2O + ZnCl_2.H_2O + 2FeCl_3 \rightarrow MnZnFe_2O_4 + 4H_2 + 5Cl_2$$

$H_2$ and $Cl_2$ were washed out during the washing and sucking process to remove impurities.

The prepared samples were annealed at 400°C, 500°C, 600°C & 700°C temperatures to study the possible crystalline structure modifications and variation in grain size, magnetic & optical properties. These as prepared samples & annealed samples at different temperature were characterized and then compared their results.

**Characterization**

The 5 different samples including as prepared, annealed at 400°C, 500°C, 600°C & 700°C are used to characterized through different characterizing mechanisms like SEM, EDX, XRD, Uv-Vis spectrometry & VSM. Scanning electron microscopy and Energy Dispersive X-ray spectroscopy were used to analyze the morphology and to conform the prepared particle's presence and their physical appearance. X-ray Diffraction spectroscopy was used to calculate the particle size, lattice parameter and d spacing. Optical properties like band gap energy was calculated by carried out results of Uv-Vis spectrometry. Vibrating Sample Magnetometer results were used to calculate the magnetic properties like Saturation Magnetization (Ms), Coercivity (Hc) & Ramanence (Mr).

# Results & Discussion

➢ X-ray Diffraction

To study the lattice parameter, phase structure and the particle size XRD analysis was carried out with Cu-Kα (λ=1.5406 Å) radiations at room temperature. The samples analysis conformed the cubic spinel structure of Particles. The experimentally generated peaks were matched with the theoretical peaks and indices. The particle sizes were calculated by The **De-bye sheerer** formula

$$D = K\lambda/\beta\cos\theta$$

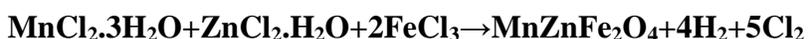

Fig#1 shows a pattern of as prepared and annealed samples with grain growth from 42 to 60 nm as with the increase of peak, the peak of as prepared and with the increase of annealing from 400˚C up to 700˚C every peak representing the grain growth and also indicating that the particles became more crystalline. Annealing make particles more refine and crystalline. The calculated particle size is shown in table#1.

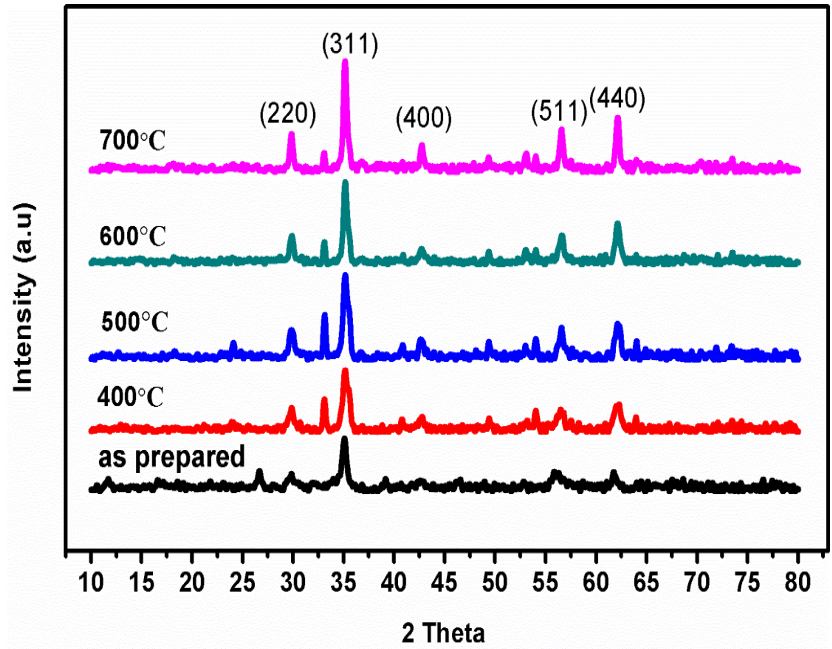

Fig#1 Indexed XRD patterns at different temperatures

Table#1 Brief description of Particle size (nm) of every peak at different annealing temperatures

| Lattices miller indices | Particle size (nm) | | | | |
|---|---|---|---|---|---|
| | As prepared | annealed at 400˚C | annealed at 500˚C | annealed at 600˚C | annealed at 700˚C |
| (220) | 34 | 26 | 17 | 20 | 41 |
| (311) | 42 | 53 | 53 | 56 | 60 |
| (400) | 17 | 18 | 21 | 21 | 36 |
| (511) | 14 | 19 | 28 | 28 | 38 |

| (440) | 16 | 24 | 14 | 19 | 38 |

The calculate particle sizes, lattice parameters and the d-spacing for max. peak is written in table #2.

Table#2 Calculated results of XRD for index (311)

| Annealing temperature | Particle size D (nm) | Lattice parameters a=b=c (nm) | d spacing (nm) |
| --- | --- | --- | --- |
| As prepared | 42 | 7.7 | 2.3 |
| annealed at 400°C | 53 | 7.7 | 2.3 |
| annealed at 500°C | 53 | 7.8 | 2.3 |
| annealed at 600°C | 56 | 7.9 | 2.3 |
| annealed at 700°C | 60 | 7.9 | 2.3 |

The graph between the particle size and the annealing show the direct relation between the annealing and the particle size as shown below.

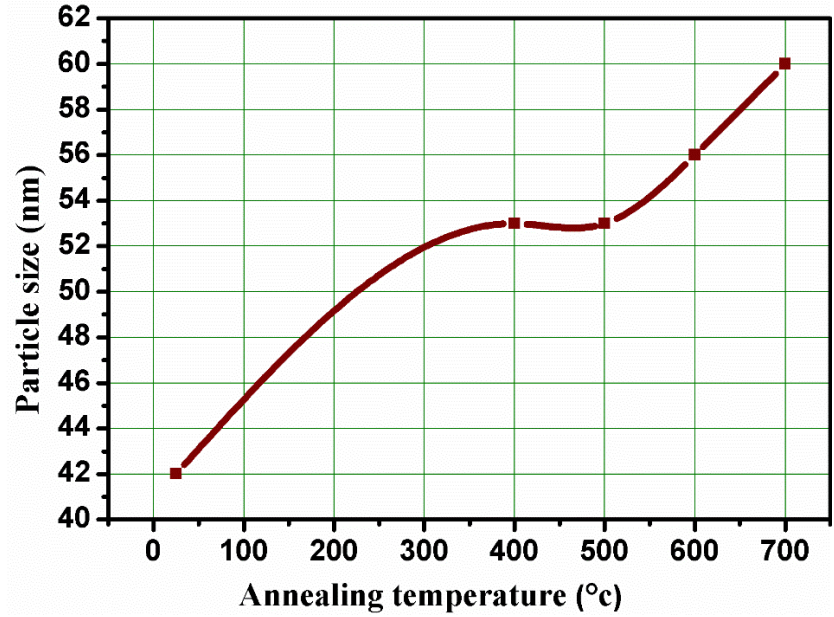

➢ SEM

The SEM technique use to observe the morphology of the Mn-Zn ferrites NPs. The SEM images of the samples as prepared and annealed at 700°C are shown below in Fig.#2(a) and 2(b).

(a)

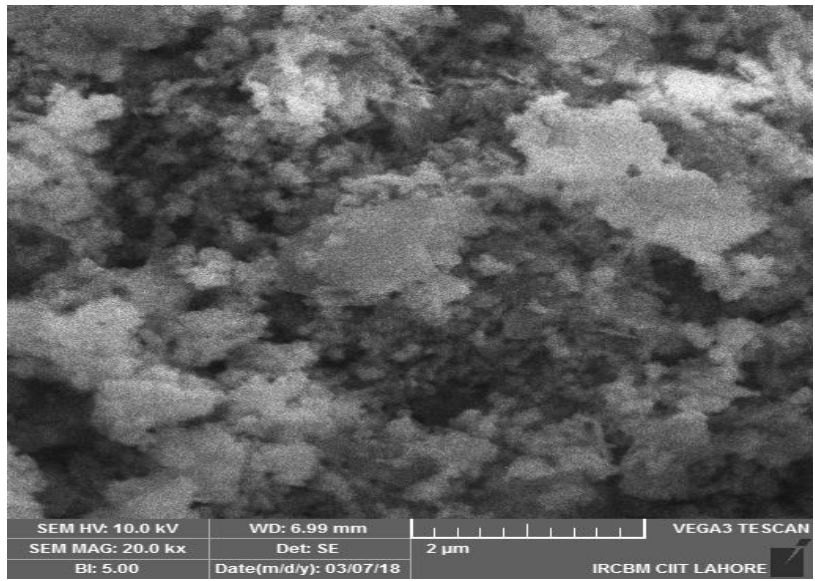

(b)

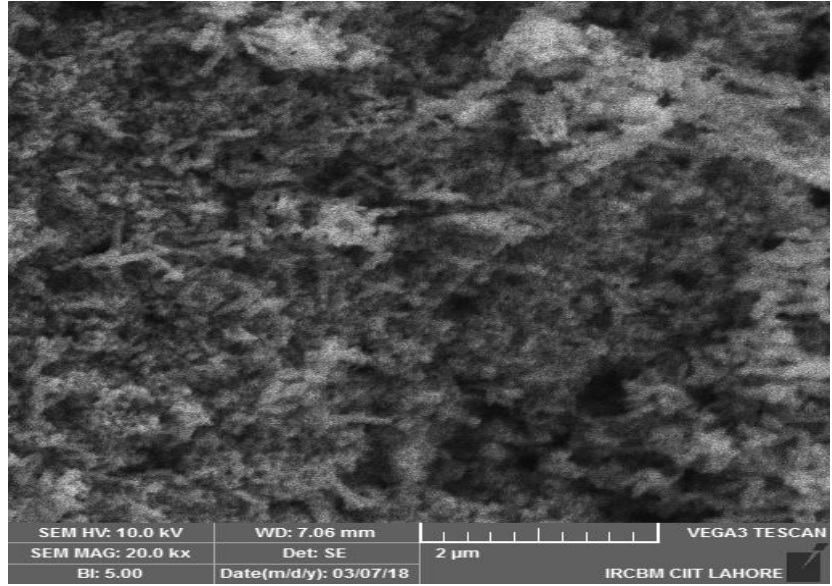

The SEM images show that the as deposited particles are more agglomerated but when the samples have been annealed at 700ºC the agglomeration reduces because of heat treatment. As prepared samples looked having flake like shapes while the samples annealed at 700ºC showed a rod like shape morphology. So, this implies that because of heat treatment there happened a longitudinal growth of the particles resulting in rod like structures at 700ºC.

- ➢ EDX

Energy dispersive graph shows the presence of Mn, Zn, Fe and $O_2$ atoms in the prepared samples. All the graphs were almost same so here we attach only one to avoid repetition.

- ➢ Uv.Vis. spectroscopy

The optical properties of Zn doped Mn ferrites are studied by using the UV-Vis spectrometry technique. The results were used to study the absorption of photon and the band gap energy. The graph between absorbance and wavelength indicates that annealing effected the absorbance. The graph showed the maximum absorbance from 364 nm to 435 nm (from near ultra to red shift) and after that absorbance decreased down gradually.

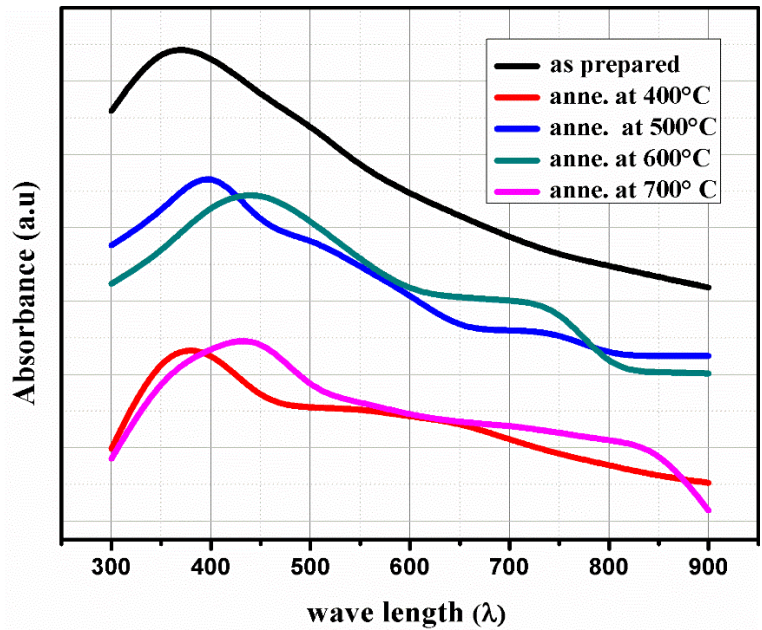

The band gap energy Eg calculated by using Tauc relation which is mentioned below.

$$h\nu\alpha = A \times (h\nu - E_g)^{1/n}$$

In this relation "α" is absorption coefficient, A is proportionality constant which is equals to 1 and Eg is band gap corresponding to a particular transition of electrons while n=1/2, 2, 3/2 and 3 characterizes the nature of transition which may be direct allowed, indirect allowed, direct forbidden and indirect forbidden transitions respectively n=1/2 showed that there is direct allowed transition occurred in Mn-Zn ferrite case.

Table#3 UV-Vis spectroscopy results

| Annealing temperature | Wavelength at max. absorbance (nm) | Band gap energy (Eg) eV |
|---|---|---|
| As prepared | 364 | 3.5 |
| 400°C | 364 | 3.25 |
| 500°C | 411 | 3.14 |
| 600°C | 435 | 3.01 |
| 700°C | 435 | 2.91 |

The calculated band gap energy of every sample is shown in Table#4. The absorbance and band gap energy decreased down as the annealing increased. The XRD results and the Uv-Vis results indicates the reverse quantization phenomena. Quantization states that "Band gap energy increases with decrease in particle size due to electron confinement at nano-scale". While in calculated results the $E_g$ decreased down with the increase of particle size.

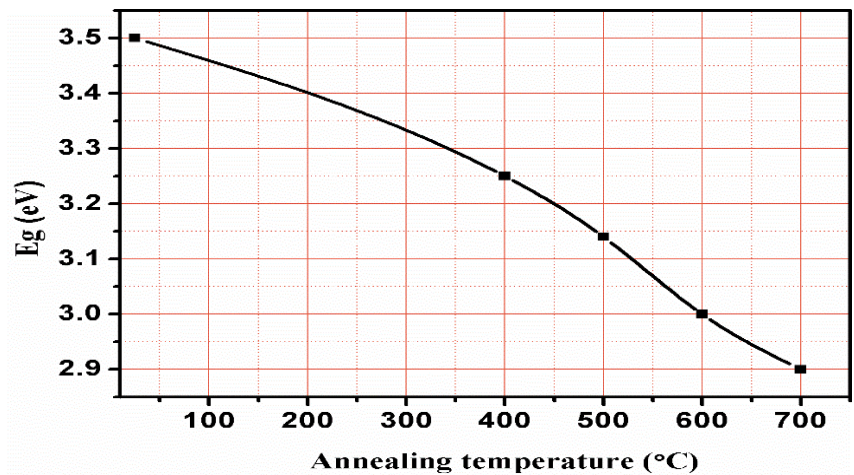

Fig#5: Graph between annealing temperature and band gap

## ➤ VSM

VSM were used to investigated the magnetic properties of the samples. The M–H loops of the annealed Mn-Zn ferrites NPs with a maximum magnetic field of ±8 kOe at room temperature. It is cleared from results that all samples have different saturation magnetization Ms values. The particles behave ferromagnetic behavior with Ms values increasing gradually from 26.5 to 65.7 emu/g. The spins at A and B lattice sites are antiparallel to each other in spinel ferrites structure.

(a)                                                                  (b)

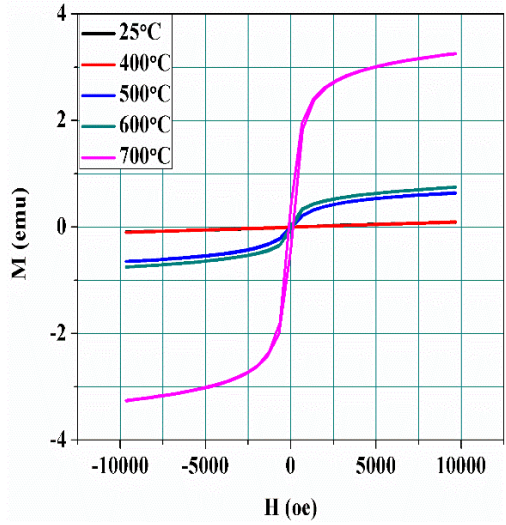 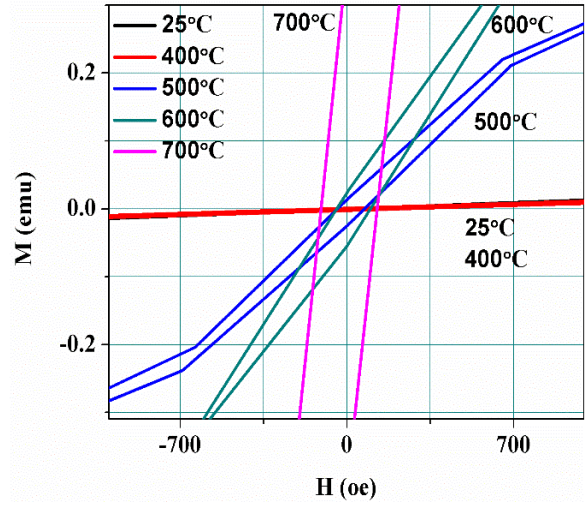

Table# VSM calculated results of Hc & Mr

| Annealing temperature (˚C) | Coercivity (Hc), (oe) | Ramanence (Mr), µ(emu) |
|---|---|---|
| As prepared | 25.640 | 3.4489 |
| 400 | 44.059 | 0.015097 |
| 500 | 53.998 | 0.001475 |
| 600 | 110.95 | 0.001475 |
| 700 | 140.04 | 367.5 |

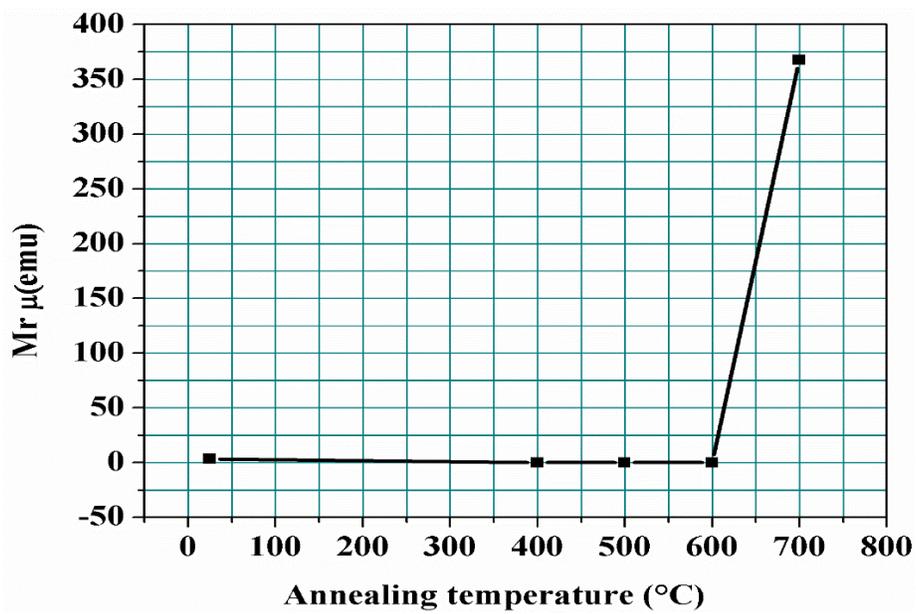

Fig.#5 Graph between annealing temperature (°C) and Retentivity (Mr) µ emu

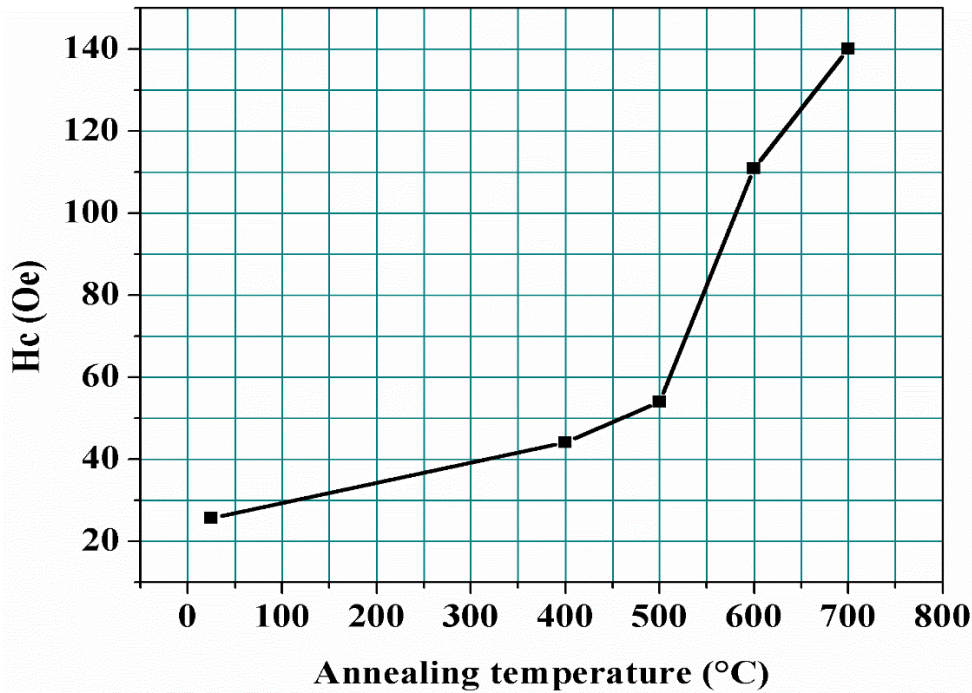

Fig.#6: Graph between annealing temperature and coercivity Hc

With high saturation magnetization Ms i.e. ±8 kOe, low coericivity and low remanence values they work as soft ferrites (easily magnetize and demagnetize as well). These materials very useful to prepare many of field applications like Ferro fluids, step up step down transformers, recording heads etc.

## Conclusions

The particles size increased with the annealing which is also confirmed by SEM results while the band gap energy decreases with the annealing which is defined as the reverse quantization process. VSM results indicates that the samples are ferromagnetic in behavior with a maximum magnetic field of ±8 kOe at room temperature with high Ms value, low coericivity and low remanence values while they work as soft ferrites.